\documentclass[conference]{IEEEtran}
\usepackage{cite}
\usepackage{graphicx}
\usepackage{tikz}
\usepackage{xcolor,colortbl}
\usepackage{amsmath}
\usepackage{multirow}
\usepackage{hhline}
\usepackage{color}
\begin{document}
\title{Acoustic Denial of Service Attacks on HDDs}

\author{
\IEEEauthorblockN{Mohammad Shahrad, Arsalan Mosenia, Liwei Song, Mung Chiang\textsuperscript{*}, David Wentzlaff, Prateek Mittal}
\IEEEauthorblockA{Princeton University~~~*Purdue University\\
Emails: \{mshahrad, arsalan, liweis, wentzlaf, pmittal\}@princeton.edu, 	*chiang@purdue.edu}
}
\markboth{Journal of \LaTeX\ Class Files,~Vol.~14, No.~8, August~2015}%
{Shell \MakeLowercase{\textit{et al.}}: Bare Demo of IEEEtran.cls for IEEE Journals}
\maketitle
\begin{abstract}
Among storage components, hard disk drives (HDDs) have become the most commonly-used type of non-volatile storage due to their recent technological advances, including, enhanced energy efficacy and significantly-improved areal density. Such advances in HDDs have made them an inevitable part of numerous computing systems, including, personal computers, closed-circuit television (CCTV) systems, medical bedside monitors, and automated teller machines (ATMs). 

Despite the widespread use of HDDs and their critical role in real-world systems, there exist only a few research studies on the security of HDDs. In particular, prior research studies have discussed how HDDs can potentially leak critical private information through acoustic or electromagnetic emanations. 

Borrowing theoretical principles from acoustics and mechanics, we propose a novel denial-of-service (DoS) attack against HDDs that exploits a physical phenomenon, known as acoustic resonance. We perform a comprehensive examination of physical characteristics of several HDDs and create acoustic signals that cause significant vibrations in HDD’s internal components. We demonstrate that such vibrations can negatively influence the performance of HDDs embedded in real-world systems. We show the feasibility of the proposed attack in two real-world case studies, namely, personal computers and CCTVs. 

\end{abstract}

\begin{IEEEkeywords}
Hard disk drives, acoustic resonance, denial-of-service, reliability, security, closed-circuit television.
\end{IEEEkeywords}
\IEEEpeerreviewmaketitle

\section{Introduction}
The rapid development of storage components, along with the miniaturization of powerful computing units, has led to an exponential increase in the number of sensing, monitoring, and computing systems, forming the modern era of computing. Among storage devices, hard disk drives (HDDs) have become the most commonly-used type of non-volatile storage in the market due to their superior advantages over other storage devices. Since their introduction in the 1950s, their energy efficacy, reliability, fault tolerance, and storage capacity have significantly improved while their sizes have decreased drastically \cite{HDD_ADVANCES_1,HDD_ADVANCES_2,HDD_ADVANCES_3}. These technological advances in HDDs, along with the ever-increasing need for storing the huge amount of data, made them one of the core components of modern computing systems. Indeed, HDDs are now an inevitable part of numerous ubiquitous systems, including, but not limited to, personal computers, cloud servers, medical bedside monitors, closed-circuit television (CCTV) systems, and automated teller machines (ATMs). 

We argue that the security of HDDs has been overlooked despite their critical role in computing systems. HDDs hold essential software components (e.g., the operating system) and various forms of sensitive information (e.g., camera footage in CCTVs), and thus, can be an appealing target for a plethora of attackers. Biedermann et al.~\cite{EM_ATTACK} have shown that HDDs can minimally leak some private information through the electromagnetic signal generated by their regular operation. Moreover, Guri et al.~\cite{Guri2017} have recently demonstrated the feasibility of using acoustic sounds created by HDDs to form a covert channel in air-gapped computers. They suggested that a malware installed on a compromised system can control the movements of the HDD's read/write arm in an attempt to generate acoustic signals, which can leak private data from the system.

We pursue an entirely different angle to the security of HDDs, motivated by the insight that computer systems heavily rely on the availability of HDDs and even a short period of unavailability/unreliability in their operations may lead to the failure of the whole system, causing serious consequences. Borrowing concepts from resonance scattering theory, we propose a novel attack against HDDs based on the acoustic resonance that can completely halt the normal operation of HDDs in real-world systems. Acoustic resonance is a well-known physical phenomenon in which an acoustic signal (generated at specific frequencies) causes vibrations in an object, potentially leading to the destruction of the object. 

Our key contributions can be summarized as follows:
\begin{enumerate}
\item To the best of our knowledge, we propose the first instance of non-contact denial of service security attacks against HDDs. 
\item We investigate how an attacker can take advantage of a well-known physical phenomenon, called acoustic resonance, to negatively affect the regular operation of HDDs, and discuss how the attacker can find the appropriate frequencies for launching the proposed attack.
\item We examine several models of state-of-the-art HDDs. As shown in Section \ref {sec:isolated_attacks}, our attack can completely halt the read/write operations of all experimented models. 
\item We further highlight the negative consequences of the proposed attack using two real-world case studies, namely, CCTV and desktop personal computer systems. We demonstrate how an attacker can disable a CCTV system by targeting its digital video recorder (DVR) device. Further, we show how the proposed attack can target a personal computer, causing a failure in its underlying OS.
\item To prove our hypothesis that the attack is caused by acoustic resonance, we disassemble an HDD and demonstrates that the attack frequencies (i.e., the frequencies at which the acoustic sound can cause a failure in the HDDs) match the resonance frequencies of HDD platters. 
\end{enumerate}
The rest of the paper is organized as follows. Section \ref{BACKGROUND} provides a short background on acoustic resonance and the structure of HDDs. Section \ref{THREAT} discusses the threat model. Section \ref{sec:isolated_attacks} describes how we perform the proposed attack on several models of HDDs and shows how an attacker can find the acoustic resonance of an HDD in a controlled environment. Section \ref{Sec:attack_demonst} shows how the proposed attack can cause serious failures in realistic scenarios, in particular, it describes the attack against two real-world systems. Section \ref{sec:attack_physics} briefly discusses how our empirical results match the outcomes predicted by the physics behind the attack. The related work is explained in Section~\ref{sec:related_work}. Finally, Section \ref{sec:conclusion} concludes the paper.


\section{Background}
\label{BACKGROUND}
In this section, we first discuss how acoustic resonance happens. We then describe the internal structure of state-of-the-art hard drives and their fundamental components.

\subsection{Acoustic Resonance}
Resonance is a physical phenomenon in which a source of vibration forces an object to oscillate with greater amplitude at specific frequencies, referred to as resonance frequencies. At resonance frequencies, even small forces generated by the source of vibration can create large-amplitude oscillations \cite{ACOUSTIC_RESONANCE}. This phenomenon has been extensively studied in mechanic as well as construction. When designing an urban infrastructure, it is essential to carefully examine its resonance frequencies.  For example, bridge designers must ensure that its natural frequencies are much greater than the expected frequency of any forces that are likely to act on the structure (for example, the wind) \cite{BRIDGE}. In this paper, we focus on a specific form of resonance, called acoustic resonance, where a source of acoustic waves creates a periodic signal at a certain frequency close to the object, causing mechanical vibrations. In particular, we aim to exploit this phenomenon to target HDDs and cause unintentional vibration in the fundamental components of HDDs.

\subsection{Structure of Hard Drives}
A typical HDD consists of two fundamental components: (i) platters, flat circular disks that are covered by a thin film of a ferromagnetic material, and (ii) read-write heads, which are positioned very close to platters. In a modern HDD, the data is stored on platters and heads are responsible for reading/writing the data from/to platters as the rotate at a very high speed (for most of the consumer-grade HDDs, platters spin at either 5,400 or 7,200 rotations per minute). Next, we briefly describe how read/write operations work.
\begin{itemize}
\item Write operation: The HDD's internal circuits control the movement of the head and the rotation of the disk, and perform writes on demand from the disk controller. For a write operation, the corresponding head (one of multiple heads that has access to the requested location of the platter) first moves to a designated area on the platter, where it can modify the magnetization of the data cells (i.e., physical spaces on the disk platter in which a single bit is stored) as they pass under it.
\item Read operation: In order to read the data stored on platters, the motor first spins up the platters, and a read-write head moves itself to the appropriate position above the platter which contains the requested data. The head can immediately detect the magnetization of the material passing under it:  changes in magnetism induce a current in the head, which is decoded by the HDD’s internal circuits and converted into a binary value. 
\end{itemize}

\section{Threat Model}
\label{THREAT}
In this section, we first describe negative consequences of acoustic resonance on HDDs and briefly mention how these consequences can affect real-world systems. We then discuss motivation of a potential attacker and our assumption about her capabilities. 
\subsection{Problem definition}
Driven by rapid advances in storage technologies, state-of-the-art HDD offer a high areal density: over 1.2 gigabits can be stored on an area of one square inch \cite{HDD_12GB}. Providing such a high areal density requires a careful design of a head positing scheme that can \textit{accurately} place read/write heads of the HDD in the appropriate position. Indeed, even a small displacement of the head leads to malfunctioning of the HDD and may even accidentally scratch the platters, causing permanent damage to the HDD. In this paper, we aim to demonstrate an active attack against HDDs in which the adversary aims to cause misplacement of internal read/write heads. As demonstrated later in Section \ref{sec:isolated_attacks}, an intentionally-generated acoustic signal can cause unwanted acoustic resonance in fundamental components of HDDs, leading to seek failures (i.e., heads cannot be precisely located at the platter location where the data will be read or written). As shown in two real-world systems (Section \ref{Sec:attack_demonst}), the proposed attack can be exploited to cause consecutive seek failures in an HDD, leading to severe failures in the whole system that relies on the HDD. For example, it can momentarily stop the normal operation of an operating system (OS) running on a personal computer or even cause it to freeze, requiring a reboot. To the best of our knowledge, the proposed attack offers the first non-contact DoS attack against HDDs. We provide a proof-of-concept of the proposed attack, shedding light on an overlooked security vulnerability of HDDs. 

\subsection{Potential attackers and their capabilities}
Due to the vital role of HDDs in numerous computing systems, we envision them to be an interesting target for a plethora of attackers. In particular, we assume that attackers may be interested in launching DoS attacks against a variety of mission-critical systems (e.g., a CCTV camera, a cloud server, or an industrial automation/monitoring system) by exploiting the susceptibilities of their HDDs. 

We assume that attackers can neither directly control nor touch the HDD. Instead, we assume the attacker can create the acoustic signal at specific frequencies to affect the availability of HDDs. We assume that the attacker can generate acoustic signals in the vicinity of the victim device, at frequencies within the audible range (2 -– 20 kHz). The attacker can either apply the signal by using an external speaker or exploit a speaker near the target. Toward this end, the attacker may potentially take advantage of remote software exploitation (for example, remotely controlling the multimedia software in a vehicle or personal device), deceive the user to play a malicious sound attached to an email or a web page \cite{TRIPPEL_1}, or embed the malicious sound in a widespread multimedia (for example, a TV advertisement). The attacker is also able to control the amplitude (consequently power) of the acoustic signal within the stated frequency range.

We do not put any limit on the ability of the attacker to study a specific targeted HDD in a controlled environment: the attacker can explore various vulnerabilities of a specific HDD or a system that has the HDD before launching the attack. Indeed, we assume that the attacker can reverse engineer a sample-computing system to extract the exact model of its HDDs and investigate the system's behavior when he creates different acoustic signals with various frequencies and amplitudes. 

\section{Isolated Acoustic Resonance Attacks}
\label{sec:isolated_attacks}
In this section, we explain how we perform acoustic DoS attacks on HDDs in isolation. By performing the acoustic attack without any barrier shielding the HDD, we maximize its exposure to sounds waves to better analyze potential vulnerabilities. Later in Section~\ref{Sec:attack_demonst}, we perform the attack on settings where hard drives are embedded in device cases.

\begin{figure}[!t]
\centering
\includegraphics[width=\linewidth]{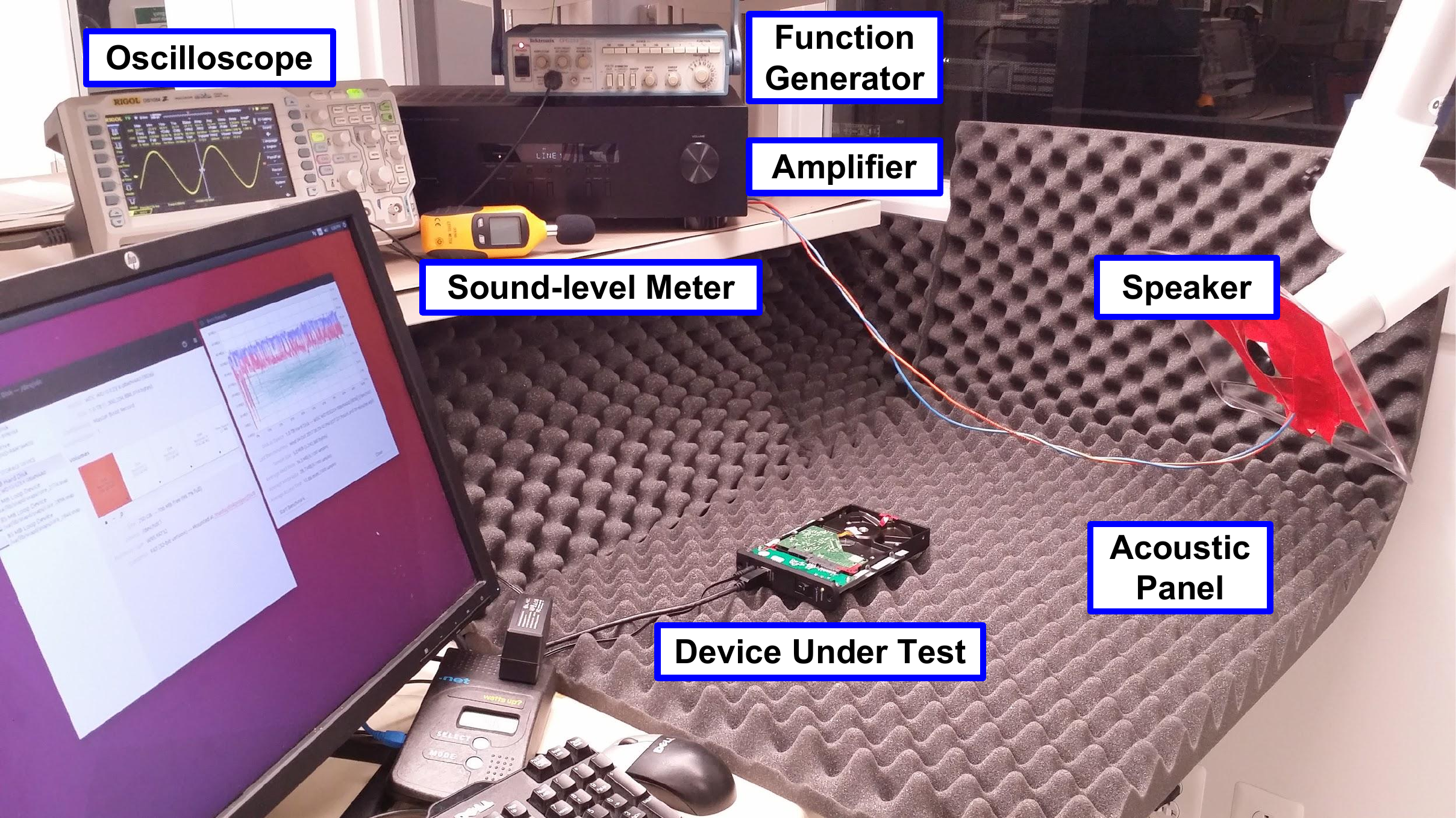}
\caption{Experimental setup for performing acoustic attacks.} 
\label{fig:setup}
\end{figure}

\begin{figure}[!t]
\centering
\includegraphics[width=\linewidth]{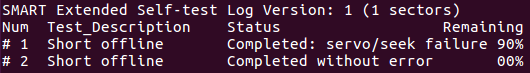}
\caption{The HDD SMART self-test fails with servo/seek failure under acoustic attack (test number 1).} 
\label{fig:smart_log}
\end{figure}

\begin{figure*}[!t]
\centering
\scalebox{0.8}{\includegraphics[width=0.98\linewidth]{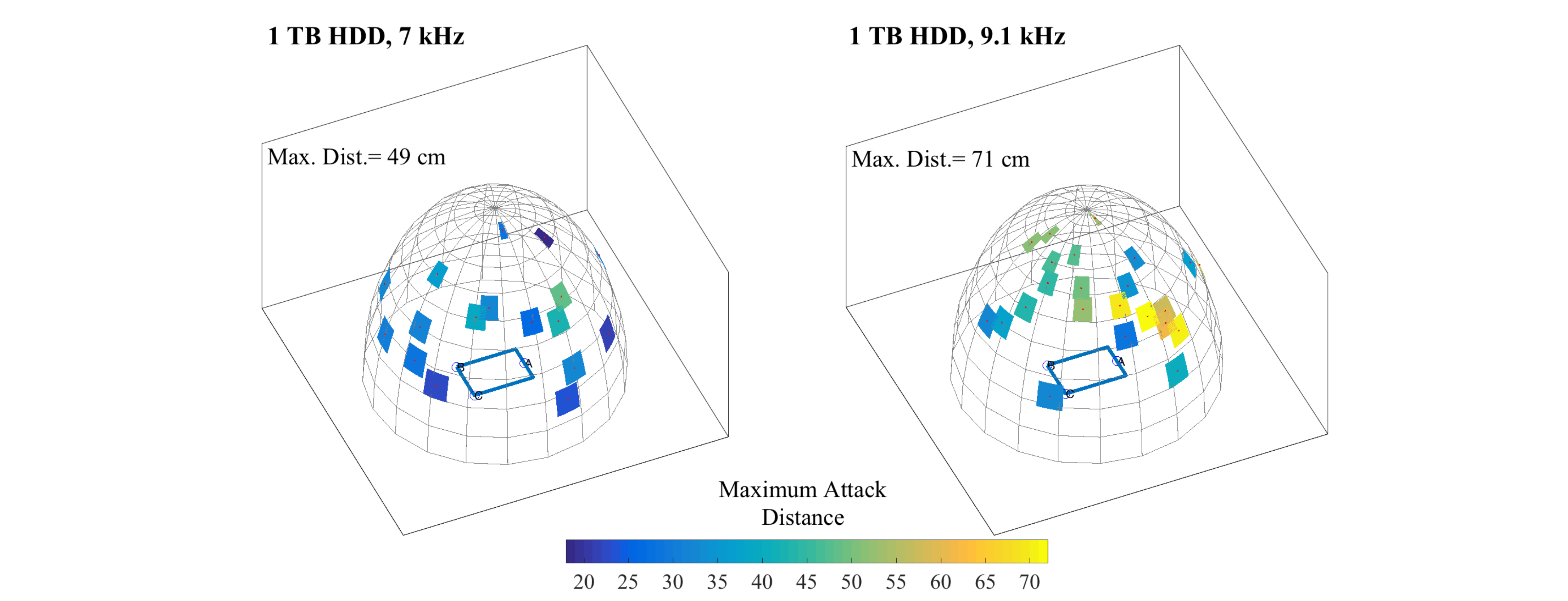}}
\scalebox{0.8}{\includegraphics[width=0.90\linewidth]{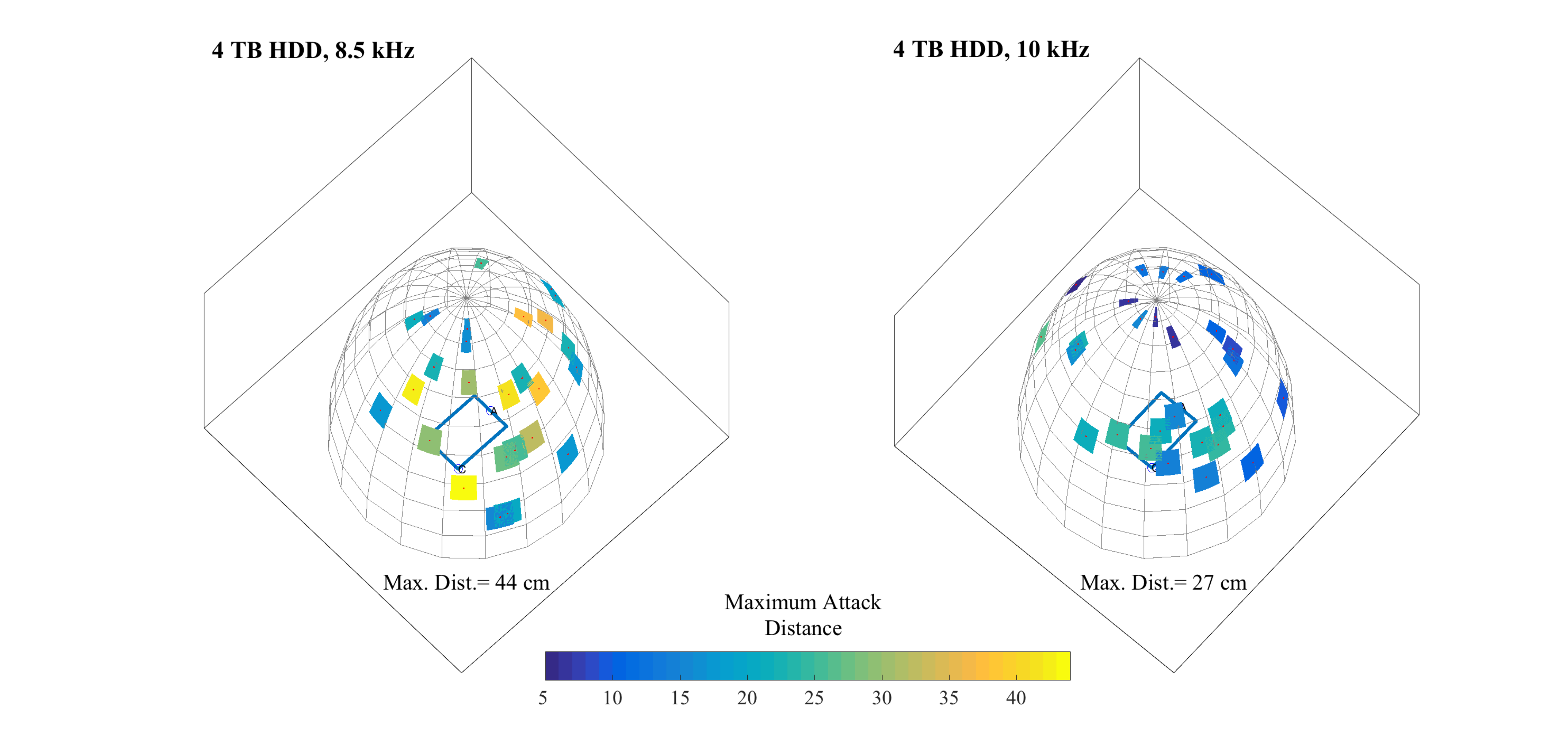}}
\caption{Maximum distance of successful acoustic attacks as a function of attack angle for a 1 TB as well as a 4 TB hard disk drive. Each HDD has been tested for two of its major resonance frequencies. }
\label{fig:1TB4TB}
\end{figure*}

\subsection{Experimental Setup}
\label{subsec:exp_setup}

Our experimental setup is depicted in Figure~\ref{fig:setup}. A function generator was used as a precise source to feed an amplifier. The output of the amplifier was then connected to a speaker and monitored by an oscilloscope. We mounted the speaker on an arm hood to be able to study the implications of strength (controlled by distance) and direction of sound waves on the feasibility of the attack. A sound-level meter was used to measure the sound level in dbA. We covered the surrounding of the device under attack with acoustic panels that absorb sound to alleviate unwanted reflections. Throughout the experiments, the operator was protected with professional earmuffs.

For testing HDDs, the target HDD was connected to a PC via a USB 3 SATA adapter. The standard read/write benchmark from the Linux Disk Utility was used to monitor the impact of sound on the performance of the disk drive. In addition, we used the Self-Monitoring, Analysis and Reporting Technology (SMART) interface through the \texttt{smartmontools} Linux package to gather detailed information on hard drive health. SMART is implemented in many modern hard drives and is widely used in HDD reliability studies~\cite{schroeder2007,li2014hard,Manousakis_2016,Mahdisoltani2017}.

\subsection{Halting Read/Write Operations through Sound}
In our first experiment, we connected different disk drives to the computer externally and exposed them to a varying sound frequency while performing the disk performance benchmark mentioned in Section \ref{subsec:exp_setup}. We then recorded frequency ranges leading to a full halt in read and write operations. During this experiment, the speaker was kept at a close distance (10 cm) with a fixed angle towards the disk under attack. We analyze the importance of the attack angle later. 

\begin{table}[!t]
\centering
\caption{Attack frequency ranges for four different HDD models.}
\begin{tabular}{ |c|c|c| }
 \hline
 \multirow{2}{11em}{\textbf{HDD Model}} & \multirow{2}{4em}{\textbf{Capacity}} & \textbf{Attack Frequency} \\
  & & \textbf{Window(s) (Hz)}\\
 \hline
 \multirow{1}{11em}{WD3200AAKS-75L9A0} & \multirow{1}{4em}{320 GB} & [2,300 - 2,510] \\
 \hline
 \multirow{3}{11em}{WD5000AAKS-75A7B0} & \multirow{3}{4em}{500 GB} & [2,240 - 2,520]\\ 
  & & [3,800 - 4,020] \\
  & & [4,725 - 5,006] \\
 \hline
 \multirow{5}{11em}{WD10EZEX-08WN4A0} & \multirow{5}{4em}{1 TB} & [2,265 - 2,281]\\ 
  & & [2,455 - 2,503] \\
  & & [6,700 - 6,845] \\
  & & [8,212 - 8,873] \\
  & & [12,839 - 12,840] \\
 \hline
 \multirow{4}{11em}{WD40EZRZ-00GXCB0} & \multirow{4}{4em}{4 TB} & [4,590 - 6,550] \\ 
  & & [7,502 - 7,900]\\
  & & [8,398 - 8,618]\\
  & & [9,420 - 10,200]\\
 \hline
\end{tabular}
\label{table:freq_windows}
\end{table}

Table~\ref{table:freq_windows} reflects the attack frequency ranges for four different HDD models with different storage capacities. While some of these attack windows are remarkably wide, some are as narrow as a few hertz. It is worth mentioning that we were not able to find any ultrasound ($>20 KHz$) attack frequencies using an ultrasound speaker. 

Following successful acoustic attacks, SMART logs of tested HDDs showed increased \textbf{\texttt{Seek\_Error\_Rate}}, an ordinarily pre-failure attribute~\cite{Manousakis_2016}. This kind of failure does not affect the data integrity but can hurt the performance. We also ran the SMART extended self-test with and without performing the acoustic attack. The test under attack failed due to \textbf{\texttt{servo/seek failure}} (Figure~\ref{fig:smart_log}) confirming the previous observation. The physics behind the attack is explained later in Section~\ref{sec:attack_physics}, but in summary, such seek errors are caused by rotational vibrations in HDD platters. This explains the higher susceptibility of denser HDDs to this attack (see Table~\ref{table:freq_windows}), since seeking smaller magnetic regions on the platter requires higher precision.

\subsection{Determining the Best Attack Angle}
Soon in our experiments, we realized that the angle of the speaker towards the hard drive has a substantial influence on the attack success. Therefore, we used the setting described in Section~\ref{subsec:exp_setup} to sample the space around the HDD under attack and measure the maximum distance for a successful attack at various angles. During all the measurements the amplitude of the sound signal fed to the speaker was fixed to 15 V. 

Figure~\ref{fig:1TB4TB} shows the spatial map of successful attack distances. The distance is color-coded and a farther distance means more vulnerability in that spherical angle. Note that only sampled angles are colored. Rectangles in this figure indicate the position of the HDD and the small $A$ marks are the side away from the SATA circuitry. While two experimented HDD models have distinct spacial attack maps, the attack frequency seems to be of great importance to conduct a successful attack. The farthest successfully performed attack was at the distance of 71 cm (92.8 dBA) for the 1 TB HDD at 9.1 kHz frequency, and 44 cm (102.6 dbA) for the 4 TB HDD at 8.5 kHz.

\section{Demonstration of the Attack in Real Settings}
\label{Sec:attack_demonst}

Hard drives are used in many electronic devices where high storage capacity is required. Using the insights from the last section, we present attacks on a CCTV Digital Video Recorder (DVR) as well as a desktop personal computer (PC) to prove vulnerability of acoustic attacks on HDDs.

\subsection{CCTV DVR}
DVRs are used to store videos recorded by the CCTV systems. Every frame of video stored on a DVR could potentially be highly crucial forensic evidence~\cite{valentine2015forensic,alshaikh2016post}. Due to the high storage capacity requirement as well as cost-effectiveness, magnetic hard drives are the prevailing storage type in DVRs. This means that our proposed acoustic DoS attack can expose a major vulnerability in almost all of CCTV systems.

\begin{figure}[!t]
\centering
\includegraphics[width=\linewidth]{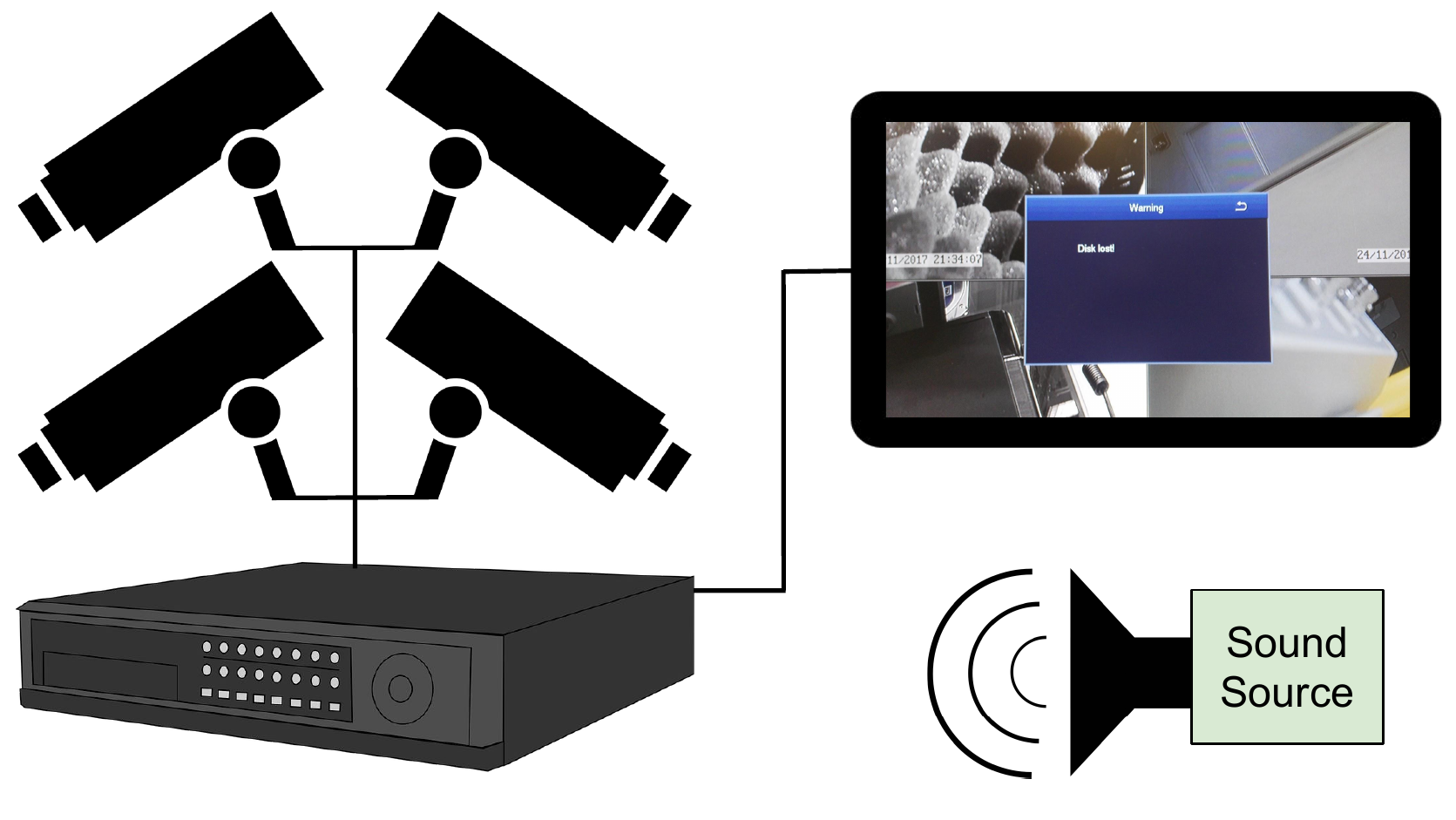}
\caption{Experimental setup for performing acoustic attacks on the digital video recorder (DVR).} 
\label{fig:dvr_setup}
\end{figure}

We evaluated this vulnerability by testing a commercial DVR (ZOSI ZR08AN/00 H.264 NDVR). We installed the 4 TB HDD that was used in the previous section in the DVR, and four digital security cameras were connected to it. We then exposed the DVR to a sound wave with fixed 8.5 kHz frequency, which is a tested major attack frequency for that HDD (see Figure~\ref{fig:1TB4TB}). We thereupon audited the monitor connected to the DVR for any anomalies. Figure~\ref{fig:dvr_setup} shows the diagram of our test setup.

\begin{figure}[!t]
\centering
\includegraphics[width=1\linewidth]{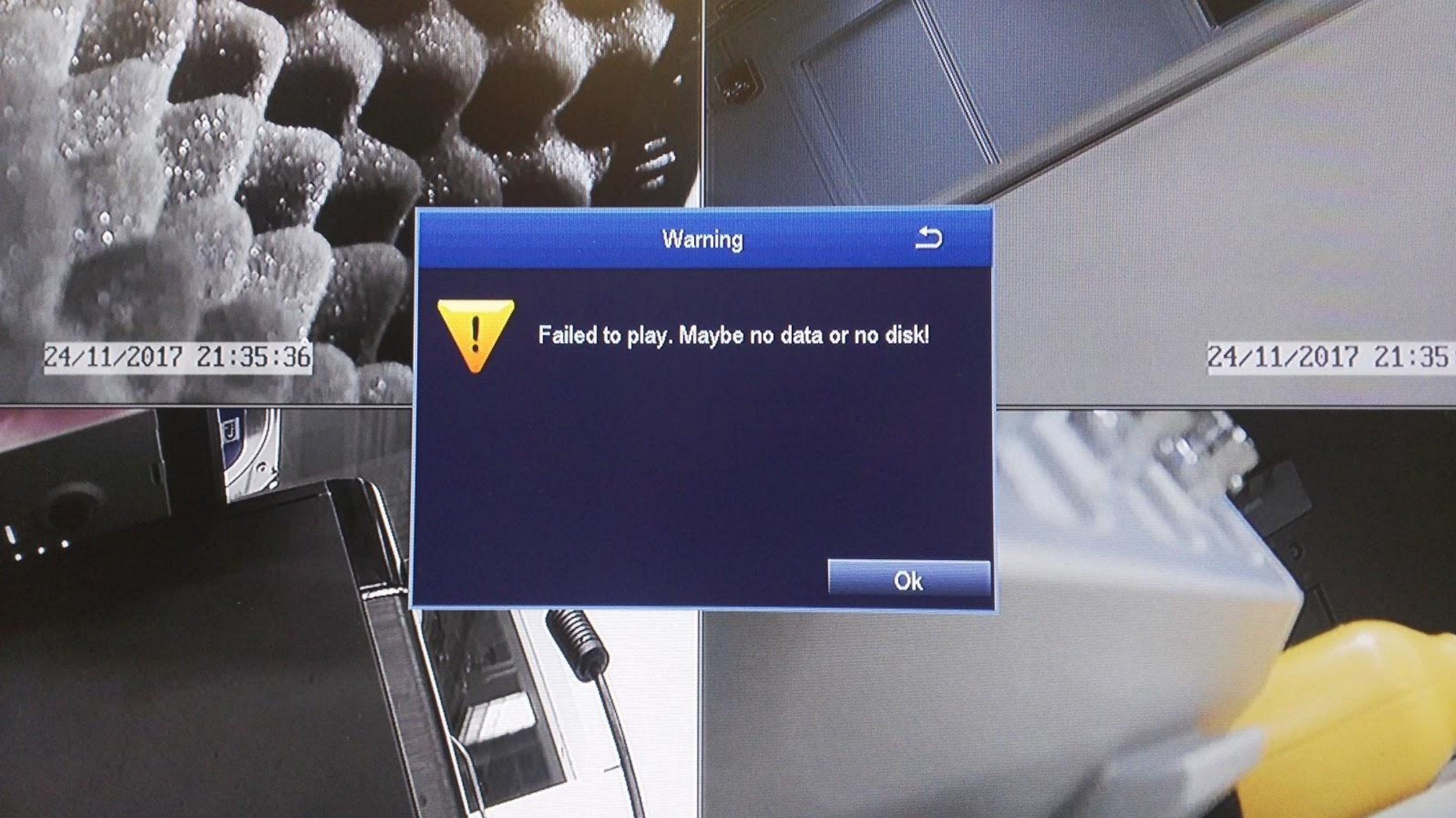}
\caption{The acoustic attack disrupting the video recording on a DVR.} 
\label{fig:failed_to_play}
\end{figure}

After around 230 seconds from starting the acoustic attack, a pop-up warning window appeared on the monitor stating \textit{``Disk lost!"}. After stopping the sound, we attempted to replay the recorded videos from four cameras and found out that recordings had been interrupted. The corresponding warning message is shown in Figure~\ref{fig:failed_to_play}. We also tried formatting the HDD using DVR's Disk Management Tool, which failed. The DVR had to be restarted to fix this issue, but the video footage was permanently lost.

We believe the reason for data loss in this scenario is memory buffer overflow. Normally, the DVR memory acts as a buffer to temporarily store video data and guarantee that no frame is lost, given an HDD's variable write speed. However, this buffer can overflow if the HDD write throughput is predominantly less than the video data generation throughput. In fact, although causing a full write termination can accelerate this attack, even a partial write slowdown can lead to a successful interruption of video recording.

It is worth mentioning that due to the blockage of sound by the DVR case, the maximum attack distance was limited to 15 cm using the same speaker setting as in the previous section. Using more powerful sound sources can increase the attack range accordingly.

\subsection{Desktop PC}

To evaluate the impact of our proposed acoustic attack on an HDD which is enclosed in a PC case, we used a desktop PC (Lenovo H520s) and installed the 1 TB HDD used in Section~\ref{sec:isolated_attacks} on it. We chose the previously tested 9.1 kHz frequency (see Figure~\ref{fig:1TB4TB}) for this experiment. We then played the fixed-frequency sound from a 25-centimeter distance towards the case's airflow opening. This has caused various kinds of malfunctions on the running PC that we report in this section.

\begin{table}[!t]
\centering
\caption{DoS symptoms of different operating systems attacked.}
\resizebox{\linewidth}{!}{
\begin{tabular}{ |c|c|c|c| } 
 \hline
 \multirow{2}{5.2em}{OS} & Attack & \multirow{2}{6.2em}{DoS Symptom} & Restart\\
  & Time & & Required \\
 \hline
 \multirow{4}{5.2em}{Windows 10} & $<5$ sec & Full file copy stoppage & N \\
 \cline{2-4}
  & \multirow{3}{2.4em}{$5$ min} & Blue screen & Y \\
  & & or & \\
  & & Error 1962: No operating system found  & Y\\
 \hline
 \multirow{2}{5.2em}{Ubuntu 16.04 LTS} & $<5$ sec & Full file copy stoppage & N\\
 \cline{2-4}
  & $1.5$ min & Unresponsive OS & Y\\
 \hline
 \multirow{2}{5.2em}{Fedora 27} & $<5$ sec & Full file copy stoppage & N\\
 \cline{2-4}
  & $2$ min & Unresponsive OS & Y\\
 \hline
\end{tabular}
}
\label{table:os_attacks}
\end{table}

\begin{figure}[!t]
\centering
\includegraphics[width=1\linewidth]{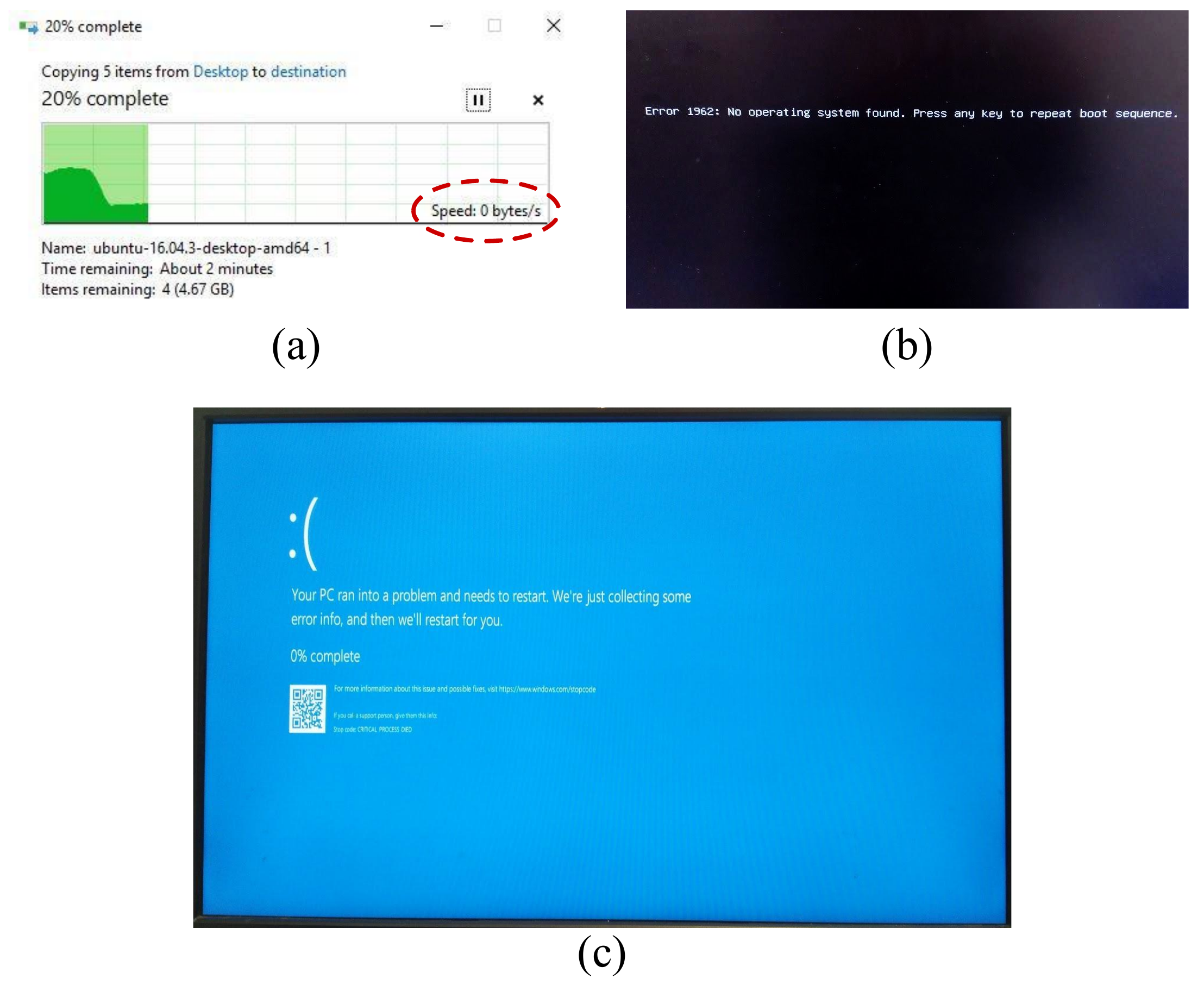}
\caption{The acoustic DoS attack can lead to different malfunctions in a desktop running Windows 10: (a) full stoppage of file copy as well as (b) black and (c) blue screen errors. } 
\label{fig:win_10_errors}
\end{figure}

We used three modern operating systems (OSs) to verify the attack. Table~\ref{table:os_attacks} summarizes different observed anomaly symptoms as a result of the acoustic attack. As seen, some of the symptoms persist after stopping the sound and require the PC to be restarted. Figure~\ref{fig:win_10_errors} shows errors when running Windows 10 as the OS.

Such behaviors are entirely expected as the hard drive is the primary storage unit containing the OS, application data, and user data. While an HDD DoS attack can directly impact disk write operations, it could also cause a critical kernel process to freeze, requiring the system to be restarted.

\section{The Physics Behind the Attack}
\label{sec:attack_physics}
\begin{figure}[!t]
\centering
\includegraphics[width=\linewidth]{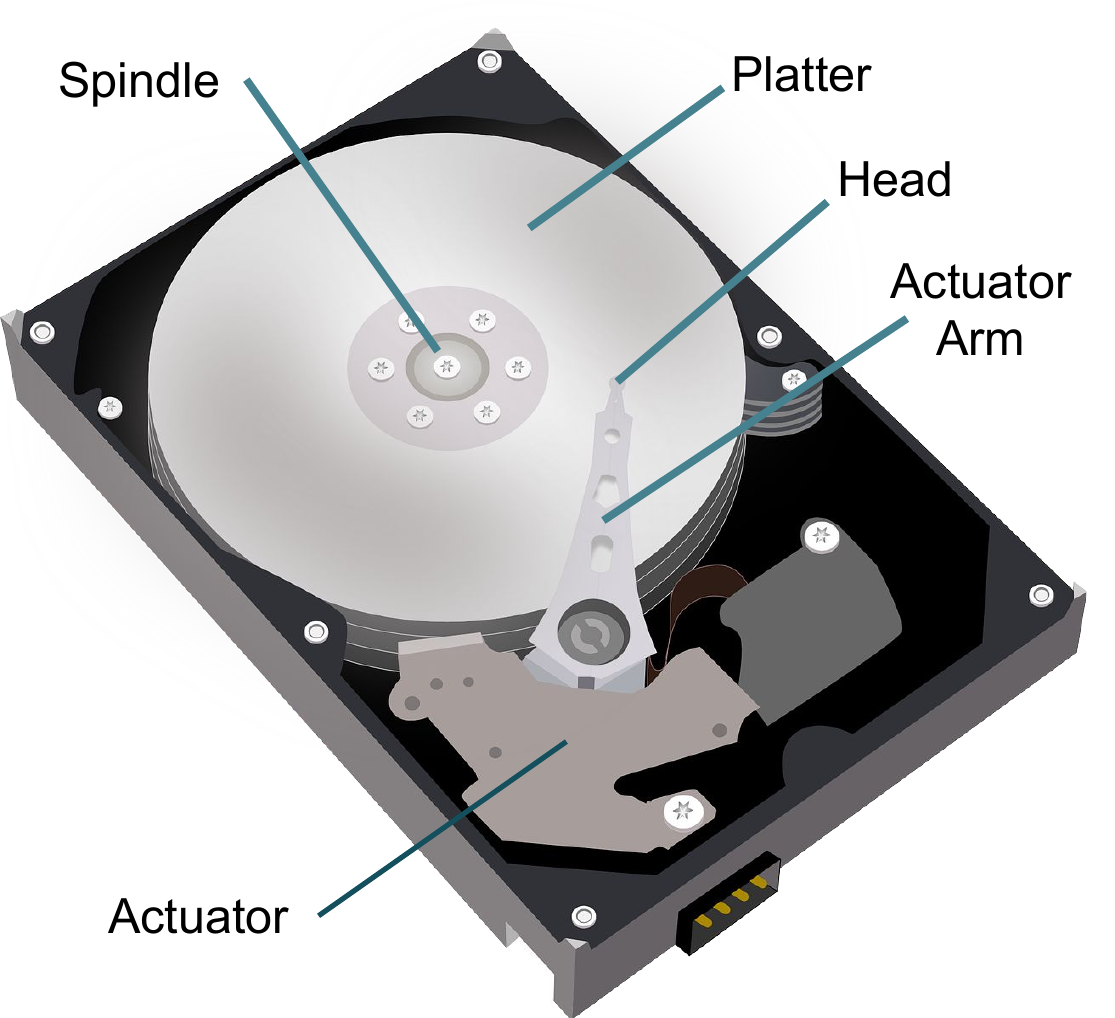}
\caption{Main moving components inside a hard disk drive.} 
\label{fig:hard_disk}
\end{figure}

Hard disk drives have one or more \textit{platters} which are covered with a magnetic coating. The information is stored as magnetic orientation in small regions on this coating layer(s). A moving head together with rotating platters enable access to any point on this surface to perform read/write operations. Figure~\ref{fig:hard_disk} shows the main moving components inside an HDD. Since a read/write operation requires precise placement of the head at specific radii of platters, any abnormal movement of these moving components can potentially lead to a failure. However, a data access happens at the granularity of a \textit{sector} -- a concentric circular track on the platter with a typical size of 512 bytes which is protected by error correcting codes (ECC). These codes have enabled hard drives to work reliably despite random failures. However, they fade if the number of failures is more than a certain threshold for a sector.

Any material object has well-defined natural oscillation eigenfrequencies, due to its constructive interference of internal and/or external surface waves~\cite{UBERALL1983307}. Resonance is caused when external waves at frequencies close to those natural eigenfrequencies are scattered by the object~\cite{flax1981theory}. We have empirically observed that disk platters make an excellent target for acoustic resonance attacks. This could be due to their relatively large surface. 

\begin{figure}[!t]
\centering
\includegraphics[width=0.95\linewidth]{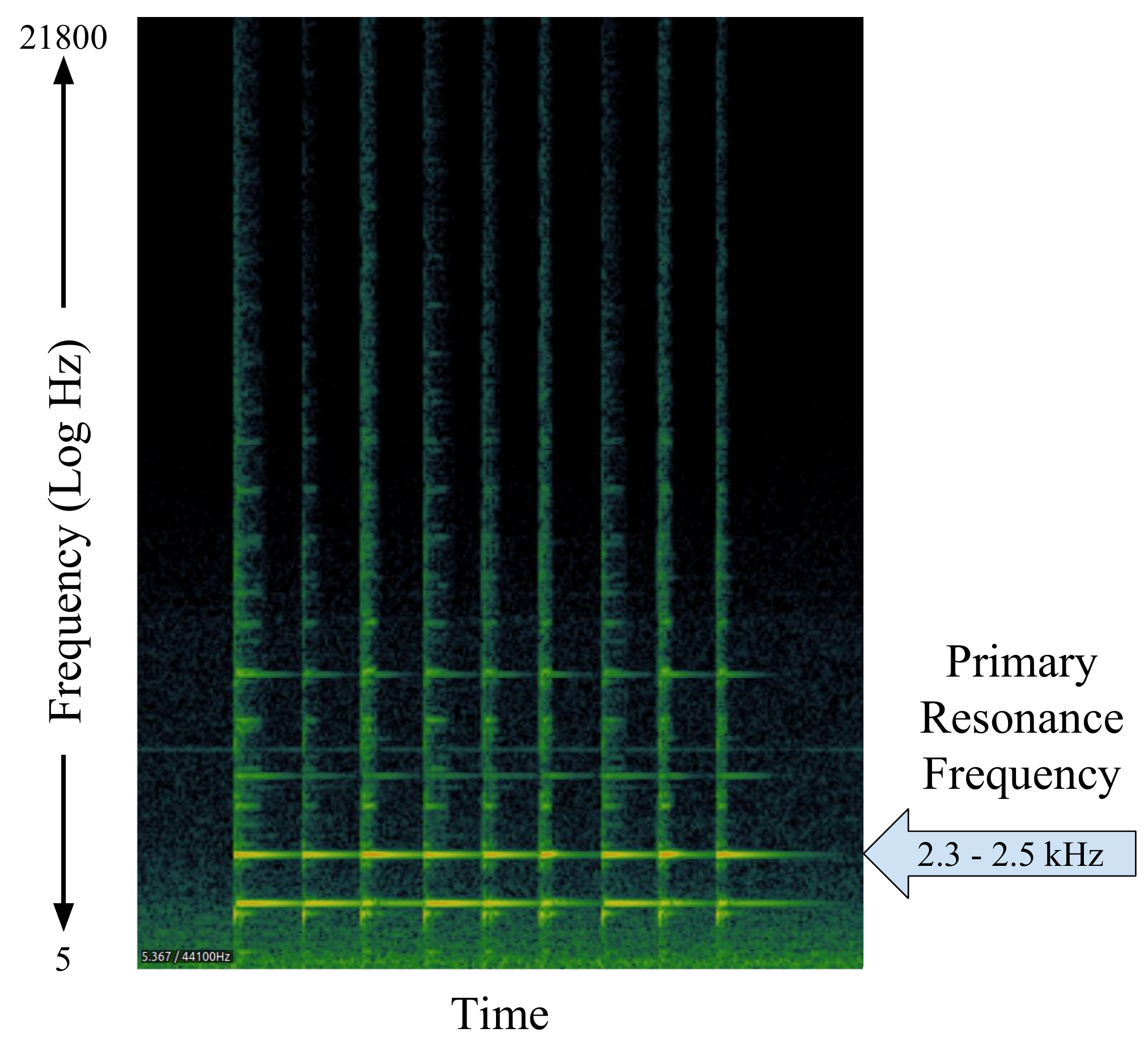}
\caption{Extracting the resonance frequencies of an HDD platter. We opened the hard drive and exerted mechanical impulses on platter, while recording the scattered sound.} 
\label{fig:platter_freqs}
\end{figure}

We have opened a hard drive to find resonance frequencies for its moving components. Using a pen to exert mechanical impulses on each component, we recorded waves scattered from it. Each component leaves a unique pattern in the frequency-domain from which the resonance frequencies become evident. The unique pattern for HDD platters of that specific hard drive (WD3200AAKS-75L9A0 in Table~\ref{table:freq_windows}) is shown in Figure~\ref{fig:platter_freqs}. For this hard drive, our observation was that frequency ranges corresponding to the successful attack are almost overlapping with the primary resonance frequency range of HDD platters.

In our proposed acoustic attack, we generate sound waves close to natural eigenfrequencies of HDD platters to cause rotational vibrations. Customarily, specific mechanisms control rotational vibrations in HDDs~\cite{suwa1999,jinzenji2001}, since they can result in read/write failures~\cite{Shah_2005}. However, it appears that existing seek control mechanisms are not able to compensate for persistent vibrations caused by the acoustic resonance. Therefore, we believe that the primary means of fortifying against this attack is to improve the seek control mechanisms of hard drives.


\section{Related Work}
\label{sec:related_work}
In this section, we discuss two categories of related research studies: previous attacks against HDDs and acoustic-based security attacks against computing systems. 

\subsection{Attacks against HDDs} 
Our proposed attack is among the first threats against HDDs. A few prior research studies have shed light on the susceptibility of storage components \cite{COLD_BOOT,EM_ATTACK}. Halderman et al. \cite{COLD_BOOT} targeted DRAMs, volatile storage components that lose their contents gradually over a period of seconds. They presented a suite of attacks, referred to as \textit{cold boot attacks} that take advantage of DRAM remanence effects in an attempt to recover cryptographic keys held in memory. They revealed a severe threat against the security of laptop users who rely on disk encryption products: they demonstrated that if an adversary steals a laptop when its encrypted disk is mounted, he can use their methodology to recover the contents. Biedermann et al. \cite{EM_ATTACK} have discussed the possibility of capturing and processing electromagnetic emanation of HDDs. They demonstrated that the magnetic field generated by the moving head of an HDD could be measured outside of the HDD to extract information about ongoing operations. In particular, they were able to detect what type of the operating system is booting up or what application is being started.
Our idea is inspired by the suggestion that environmental conditions, such as a high level of humidity or ambient noise, may negatively affect the reliability of HDDs \cite{ENV_0,ENV_1, Manousakis_2016}. In particular, in a blog post, Kruszelnicki \cite{ENV_0} suggested that loud environmental noise can degrade the performance of a hard drive: he discussed that shouting at spinning HDDs located in data centers may significantly drop their performance. The post states that this observation was initially made back in 2008 by Brendan Gregg, an Australian computer engineer (Brendan Gregg uploaded a demo on YouTube \cite{ENV_1}). 
Finally, a recent demo has demonstrated the HDD acoustic resonance attack at a fixed frequency (130Hz) on a Linux machine~\cite{Resonance_YT}. We were not able to conduct any successful attack or performance degradation for our tested HDDs in that frequency.

\subsection{Acoustic attacks against computing systems} 
A few recent studies have discussed how the presence of acoustic noise can negatively degrade the performance of input devices, in particular, sensors and microphones, used in real-world systems. Dean et al. \cite{DEAN_1,DEAN_2} have examined how the performance of Micro-Electro-Mechanical Systems (MEMS) gyroscopes may be negatively affected by environmental acoustic noise. Son et al. \cite{Son_2015} proposed an attack against MEMS gyroscopes embedded in drones and demonstrated that an attacker could incapacitate drones using intentional sound noise. Trippel et al. \cite{TRIPPEL_1} investigated how analog acoustic injection attacks can damage the digital integrity of widely-used MEMS accelerometer. They showed that spoofing accelerometer data with intentional acoustic interference provides a pathway for adversaries to deliver arbitrary digital values to embedded systems, which trust the integrity of sensor readings. Carlini et al. \cite{CARLINI_1} and Vaidya \cite{VAIDYA_1} have shown that obfuscated (hidden in another sound in the audible range) voice commands can be interpreted by speech recognition systems. In two independent studies, Zhang et al. \cite{ZHANG_1} and Song et al. \cite{SONG_1} have shown the vulnerability of microphones against inaudible acoustic commands. They exploited the nonlinearity of the microphone circuits and created inaudible voice commands that can be successfully picked up, demodulated, recovered, and interpreted by the speech recognition systems, such as Cortana, Alexa, and Siri. In this paper, we examined how intentionally-created acoustic signals can degrade the performance of a fundamental component of computing and embedded systems, namely HDDs. 


\section{Conclusion} 
\label{sec:conclusion}
Among storage components, HDDs have become the most commonly-used type of non-volatile storage due to their enhanced energy efficiency, significantly-improved areal density, and low cost. These have made them an inevitable part of numerous computing systems, including, personal computers, bedside monitors, CCTVs, and ATMs. 
Borrowing concepts from acoustics and mechanics, we presented the first instance of non-contact DoS attack against HDDs. The attack relies on a physical phenomenon, known as acoustic resonance, in which a sound wave forces an object to vibrate with a higher amplitude at specific frequencies. Our attack is motivated by the insight that real-world systems heavily rely on the availability of HDDs and inspired by the observation that acoustic resonance can cause vibrations in an object. 
We conducted a thorough examination of physical characteristics of fundamental components of HDDs (read/write head and platters) and created acoustic signals that cause significant vibration in HDD platters. These oscillations halt write/read operation of embedded HDDs, leading to severe security concerns in mission-critical systems. We demonstrated the feasibility of the attack in two real-world case studies, namely, CCTVs, and personal computers. Our proof-of-concept demonstrations shed light on a new security threat against computing systems, paving the way for further exploring overlooked vulnerabilities of HDDs.

\bibliography{references.bib}{}

\begin{thebibliography}{10}
\providecommand{\url}[1]{#1}
\csname url@samestyle\endcsname
\providecommand{\newblock}{\relax}
\providecommand{\bibinfo}[2]{#2}
\providecommand{\BIBentrySTDinterwordspacing}{\spaceskip=0pt\relax}
\providecommand{\BIBentryALTinterwordstretchfactor}{4}
\providecommand{\BIBentryALTinterwordspacing}{\spaceskip=\fontdimen2\font plus
\BIBentryALTinterwordstretchfactor\fontdimen3\font minus
  \fontdimen4\font\relax}
\providecommand{\BIBforeignlanguage}[2]{{%
\expandafter\ifx\csname l@#1\endcsname\relax
\typeout{** WARNING: IEEEtran.bst: No hyphenation pattern has been}%
\typeout{** loaded for the language `#1'. Using the pattern for}%
\typeout{** the default language instead.}%
\else
\language=\csname l@#1\endcsname
\fi
#2}}
\providecommand{\BIBdecl}{\relax}
\BIBdecl

\bibitem{HDD_ADVANCES_1}
M.~Re, ``Tech talk on {HDD} areal density,'' \url{
  https://www.seagate.com/www-content/investors/\_shared/docs/tech-talk-mark-re-20150825.pdf},
  accessed: 2017-12-10.

\bibitem{HDD_ADVANCES_2}
M.~Igami and K.~Uetake, ``Mergers, innovation, and entry-exit dynamics:
  Consolidation of the hard disk drive industry, 1996-2015,'' 2016.

\bibitem{HDD_ADVANCES_3}
T.~Yamaguchi, M.~Hirata, and J.~C.~K. Pang, \emph{High-speed precision motion
  control}.\hskip 1em plus 0.5em minus 0.4em\relax CRC press, 2017.

\bibitem{EM_ATTACK}
S.~Biedermann, S.~Katzenbeisser, and J.~Szefer, ``Hard drive side-channel
  attacks using smartphone magnetic field sensors,'' in \emph{International
  Conference on Financial Cryptography and Data Security}.\hskip 1em plus 0.5em
  minus 0.4em\relax Springer, 2015, pp. 489--496.

\bibitem{Guri2017}
M.~Guri, Y.~Solewicz, A.~Daidakulov, and Y.~Elovici, \emph{Acoustic Data
  Exfiltration from Speakerless Air-Gapped Computers via Covert Hard-Drive
  Noise (`DiskFiltration')}.\hskip 1em plus 0.5em minus 0.4em\relax Cham:
  Springer International Publishing, 2017, pp. 98--115.

\bibitem{ACOUSTIC_RESONANCE}
R.~H. Randall, \emph{An introduction to acoustics}.\hskip 1em plus 0.5em minus
  0.4em\relax Courier Corporation, 2012.

\bibitem{BRIDGE}
K.~Y. Billah and R.~H. Scanlan, ``Resonance, {Tacoma Narrows} bridge failure,
  and undergraduate physics textbooks,'' \emph{American J. Physics}, vol.~59,
  no.~2, pp. 118--124, 1991.

\bibitem{HDD_12GB}
M.~Re, ``Hackers can now steal data by listening to the sound of a computer's
  hard drive,''
  \url{https://www.forbes.com/sites/tomcoughlin/2015/06/28/progress-in-hdd-areal-density/\#4f4554a61671},
  accessed: 2017-12-10.

\bibitem{TRIPPEL_1}
T.~Trippel, O.~Weisse, W.~Xu, P.~Honeyman, and K.~Fu, ``{WALNUT}: Waging doubt
  on the integrity of mems accelerometers with acoustic injection attacks,'' in
  \emph{In Proceedings of the 2nd IEEE European Symposium on Security and
  Privacy (EuroS\&P 2017). To appear}.

\bibitem{schroeder2007}
B.~Schroeder and G.~A. Gibson, ``Disk failures in the real world: What does an
  mttf of 1, 000, 000 hours mean to you?'' in \emph{FAST}, vol.~7, no.~1, 2007,
  pp. 1--16.

\bibitem{li2014hard}
J.~Li, X.~Ji, Y.~Jia, B.~Zhu, G.~Wang, Z.~Li, and X.~Liu, ``Hard drive failure
  prediction using classification and regression trees,'' in \emph{Dependable
  Systems and Networks (DSN), 2014 44th Annual IEEE/IFIP International
  Conference on}.\hskip 1em plus 0.5em minus 0.4em\relax IEEE, 2014, pp.
  383--394.

\bibitem{Manousakis_2016}
I.~Manousakis, S.~Sankar, G.~McKnight, T.~D. Nguyen, and R.~Bianchini,
  ``Environmental conditions and disk reliability in free-cooled datacenters,''
  in \emph{14th {USENIX} Conference on File and Storage Technologies ({FAST}
  16)}.\hskip 1em plus 0.5em minus 0.4em\relax Santa Clara, CA: {USENIX}
  Association, 2016, pp. 53--65.

\bibitem{Mahdisoltani2017}
F.~Mahdisoltani, I.~Stefanovici, and B.~Schroeder, ``Proactive error prediction
  to improve storage system reliability,'' in \emph{2017 {USENIX} Annual
  Technical Conference ({USENIX} {ATC} 17)}.\hskip 1em plus 0.5em minus
  0.4em\relax Santa Clara, CA: {USENIX} Association, 2017, pp. 391--402.

\bibitem{valentine2015forensic}
T.~Valentine and J.~P. Davis, \emph{Forensic facial identification: Theory and
  practice of identification from eyewitnesses, composites and CCTV}.\hskip 1em
  plus 0.5em minus 0.4em\relax John Wiley \& Sons, 2015.

\bibitem{alshaikh2016post}
A.~AlShaikh and M.~Sedky, ``Post incident analysis framework for automated
  video forensic investigation,'' \emph{International Journal of Computer
  Applications}, vol. 135, no.~12, pp. 1--7, 2016.

\bibitem{UBERALL1983307}
H.~Überall, P.~Moser, J.~Murphy, A.~Nagl, G.~Igiri, J.~Subrahmanyam,
  G.~Gaunard, D.~Brill, P.~Delsanto, J.~Alemar, and E.~Rosario,
  ``Electromagnetic and acoustic resonance scattering theory,'' \emph{Wave
  Motion}, vol.~5, no.~4, pp. 307 -- 329, 1983.

\bibitem{flax1981theory}
L.~Flax, G.~C. Gaunaurd, and H.~Uberall, ``Theory of resonance scattering,''
  \emph{Physical acoustics}, vol.~15, pp. 191--294, 1981.

\bibitem{suwa1999}
M.~Suwa and K.~Aruga, ``Evaluation system for residual vibration from hdd
  mounting mechanism,'' \emph{IEEE transactions on magnetics}, vol.~35, no.~2,
  pp. 868--873, 1999.

\bibitem{jinzenji2001}
A.~Jinzenji, T.~Sasamoto, K.~Aikawa, S.~Yoshida, and K.~Aruga, ``Acceleration
  feedforward control against rotational disturbance in hard disk drives,''
  \emph{IEEE Transactions on Magnetics}, vol.~37, no.~2, pp. 888--893, 2001.

\bibitem{Shah_2005}
S.~Shah and J.~G. Elerath, ``Reliability analysis of disk drive failure
  mechanisms,'' in \emph{Annual Reliability and Maintainability Symposium,
  2005. Proceedings.}, Jan 2005, pp. 226--231.

\bibitem{COLD_BOOT}
J.~A. Halderman, S.~D. Schoen, N.~Heninger, W.~Clarkson, W.~Paul, J.~A.
  Calandrino, A.~J. Feldman, J.~Appelbaum, and E.~W. Felten, ``Lest we
  remember: cold-boot attacks on encryption keys,'' \emph{Communications of the
  ACM}, vol.~52, no.~5, pp. 91--98, 2009.

\bibitem{ENV_0}
``Loud sounds can kill computer hard drives,'' \url{
  http://www.abc.net.au/radionational/programs/greatmomentsinscience/loud-sounds-can-kill-computer-hard-drives/7938388
  }, accessed: 2017-12-10.

\bibitem{ENV_1}
``{Shouting in the Datacenter},''
  \url{https://www.youtube.com/watch?v=tDacjrSCeq4}, accessed: 2017-12-10.

\bibitem{Resonance_YT}
``{Resonance attack against HDD~},''
  \url{https://www.youtube.com/watch?v=8DdqTz3CW5Y}, accessed: 2017-11-19.

\bibitem{DEAN_1}
R.~N. Dean, S.~T. Castro, G.~T. Flowers, G.~Roth, A.~Ahmed, A.~S. Hodel, B.~E.
  Grantham, D.~A. Bittle, and J.~P. Brunsch, ``A characterization of the
  performance of a mems gyroscope in acoustically harsh environments,''
  \emph{IEEE Transactions on Industrial Electronics}, vol.~58, no.~7, pp.
  2591--2596, 2011.

\bibitem{DEAN_2}
R.~N. Dean, G.~T. Flowers, A.~S. Hodel, G.~Roth, S.~Castro, R.~Zhou,
  A.~Moreira, A.~Ahmed, R.~Rifki, B.~E. Grantham \emph{et~al.}, ``On the
  degradation of mems gyroscope performance in the presence of high power
  acoustic noise,'' in \emph{Industrial Electronics, 2007. ISIE 2007. IEEE
  International Symposium on}.\hskip 1em plus 0.5em minus 0.4em\relax IEEE,
  2007, pp. 1435--1440.

\bibitem{Son_2015}
Y.~Son, H.~Shin, D.~Kim, Y.~Park, J.~Noh, K.~Choi, J.~Choi, and Y.~Kim,
  ``Rocking drones with intentional sound noise on gyroscopic sensors,'' in
  \emph{24th {USENIX} Security Symposium ({USENIX} Security 15)}.\hskip 1em
  plus 0.5em minus 0.4em\relax Washington, D.C.: {USENIX} Association, 2015,
  pp. 881--896.

\bibitem{CARLINI_1}
N.~Carlini, P.~Mishra, T.~Vaidya, Y.~Zhang, M.~Sherr, C.~Shields, D.~Wagner,
  and W.~Zhou, ``Hidden voice commands.'' in \emph{USENIX Security Symposium},
  2016, pp. 513--530.

\bibitem{VAIDYA_1}
T.~Vaidya, ``Cocaine {Noodles}: exploiting the gap between human and machine
  speech recognition,'' \emph{Presented at WOOT}, vol.~15, pp. 10--11, 2015.

\bibitem{ZHANG_1}
G.~Zhang, C.~Yan, X.~Ji, T.~Zhang, T.~Zhang, and W.~Xu, ``{DolphinAttack}:
  Inaudible voice commands,'' \emph{arXiv preprint arXiv:1708.09537}, 2017.

\bibitem{SONG_1}
L.~Song and P.~Mittal, ``Inaudible voice commands,'' \emph{arXiv preprint
  arXiv:1708.07238}, 2017.

\end{thebibliography}
\bibliographystyle{IEEEtran}


\end{document}